# Resonant Processing of Instrumental Sound Controlled by Spatial Position

**Camille Goudeseune**
Integrated Systems Laboratory,
University of Illinois at
Urbana-Champaign
Urbana, IL 61801 USA
+1 217.244.9662
cog@uiuc.edu

**Guy Garnett**
Integrated Systems Laboratory,
University of Illinois at
Urbana-Champaign
Urbana, IL 61801 USA
+1 217.244.0592
garnett@uiuc.edu

**Timothy Johnson**
School of Music,
University of Illinois at
Urbana-Champaign
Urbana, IL 61801 USA
+1 217.333.1712
tejohnso@uiuc.edu

**ABSTRACT**
We present an acoustic musical instrument played through a resonance model of another sound. The resonance model is controlled in real time as part of the composite instrument. Our implementation uses an electric violin, whose spatial position modifies filter parameters of the resonance model. Simplicial interpolation defines the mapping from spatial position to filter parameters. With some effort, pitch tracking can also control the filter parameters. The individual technologies – motion tracking, pitch tracking, resonance models – are easily adapted to other instruments.

**INTRODUCTION**
The *eviolin* is an electric violin augmented in several ways. A Linux PC tracks the violin's pitch, amplitude, spectral brightness, and position/orientation of violin body and bow. The signal picked up at the violin's bridge is processed through a filter network running on a Macintosh, allowing for a range of processing from subtle to extreme with several degrees of freedom. Sound is rendered with a consumer-electronics approximation to a spherical loudspeaker, to better blend with other instruments in an ensemble.

**SOUND PROCESSING WITH RESONANCE MODELS**
The filter network is implemented in Opcode Max using CNMAT's resonance model object [3]. It essentially lets the performer play the violin "through" a tubular bell, double bass body, piano string, or artificial resonator. Each resonance model consists of a set of resonances, or filters. Each filter has a gain, center frequency, bandwidth, and decay time. Bandwidth is coupled to decay time: wide bandwidths have short decay times and vice versa. Tracking data from the Linux PC modulates the filters in real time. This communication between the Linux PC and the Macintosh uses the OpenSound Control (OSC) network protocol [7].

**Controlling gain**
It is essential to control the overall gain from the signal at the bridge of the violin to the loudspeaker's output. The instrument feels wrong if it gets suddenly louder without a corresponding change of bowing, or if it refuses to get louder no matter how strongly the bow is played. Gain control is tricky because the filters sometimes block all frequencies present in the bridge signal, and at other times excessively boost them. In early experimentation we found it useful to add to the resonance model's input some white noise, scaled by the violin's own amplitude.

The simplest gain control comes from scaling the output signal to follow the bridge signal. The bridge signal goes to MSP's avg~ object [2], which measures the average signal amplitude 130 times per second. (130 Hz is the lowest frequency of this five-string violin; we need a time window at least as long as a full period of this lowest frequency to accurately measure average amplitude). This signal level is sent to MSP's normalize~ object [2] to scale the output of the filter bank.

As is well known, resonances with high Q cause problems with gain control: the output signal becomes suddenly louder if one of the partials of the input signal lines up with such a resonance (either because the violinist changed pitch, or because the resonance changed due to movement of the violin). We therefore avoid such narrow resonances.

Some resonance models come from pitched sources like tubular bells or notes played on a piano. We have tried scaling such resonance models to match the violin's tracked pitch. At present, pitchtracking latency sometimes causes objectionable artifacts as the filters swoop to new values after a transition between two notes.

**TRACKING PITCH AND SPATIAL POSITION**

Tracking of pitch, amplitude, and spectral centroid is done with a modified MSP fiddle~ object [5] running in the sound synthesis package VSS [1]. (The spectral centroid is computed from the energy in individual partials, once pitch tracking has determined the sound's fundamental frequency.) Motion tracking is done with the Ascension SpacePad, a device which measures the position of sensors relative to an antenna emitting a time-varying magnetic field. One sensor is fastened to the back of the violin body; the other is sewn onto a fingerless glove worn on the violinist's bow hand (figure 1). The fiddle~ object therefore measures 3 dimensions, while the SpacePad measures 12 more (two sensors each measuring x, y, z, yaw, pitch, and roll).

VSS can be used to synthesize sound directly, but in this application it just sends the tracking data to the Macintosh, via a protocol implemented in OSC. Messages which the Macintosh sends include commands to open and close a connection, and commands to request certain kinds of data at certain rates. For example, the command "/ViolinControl/Param/Amplitude 30" causes VSS to report the current amplitude of the bridge signal every 30 msec; "/ViolinControl/Param/Z 100" reports the current Z-coordinate every 100 msec. A value of zero in such a command causes VSS to report that datum as fast as possible, typically every 3 to 10 msec; a negative value halts the reporting. This reporting is done with OSC messages of the same name: VSS sends "/ViolinControl/Param/Z 0.3" back to the Macintosh to indicate a Z-value of 0.3. In essence, the Linux PC is an "eviolin server", analogous to a web server presenting real-time data; the Macintosh is a client which connects to this server.

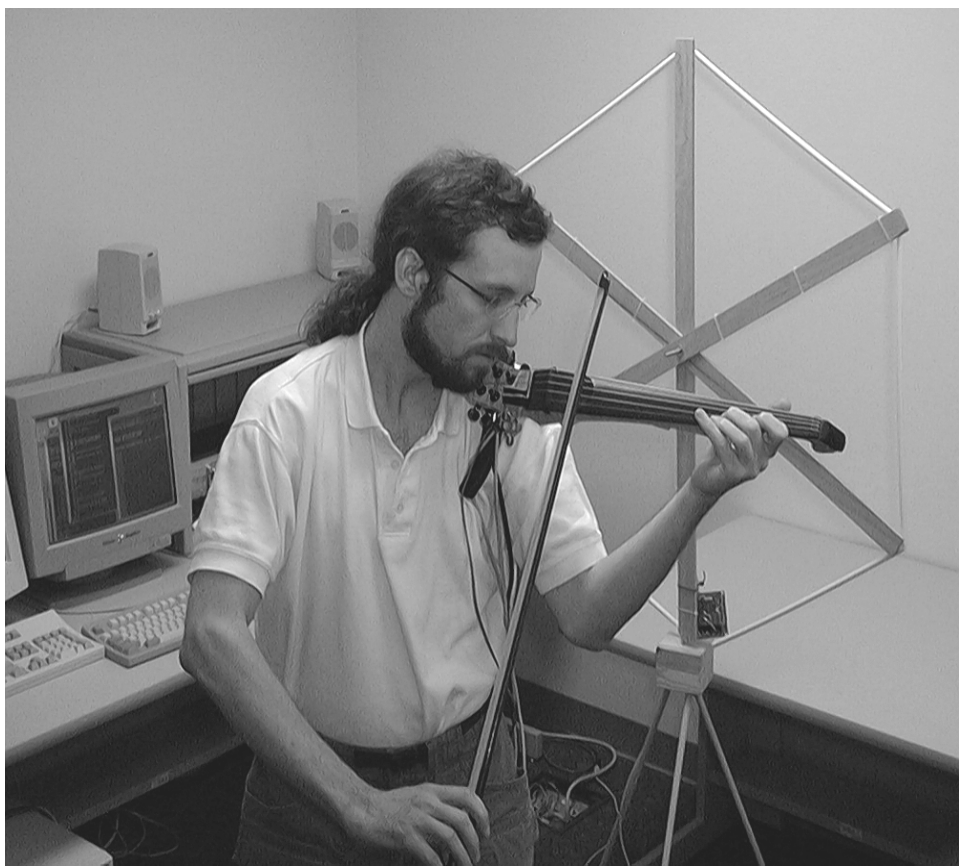

Figure 1. The eviolin and motion-tracking antenna. Photograph by Bill Sherman.

**MAPPING TRACKED DATA TO FILTER PARAMETERS**

The latitude and longitude of the violin is mapped to timbre roughly according to perceptual dimensions: spectral brightness (relative prominence of higher partials) varies with latitude, spectral richness (number of partials) with longitude. Though a third timbral dimension could be directly controlled with altitude, we have found extended playing at nonstandard altitudes to be fatiguing. Instead, we use altitude to control other things – a toggle switch such as an octave switch, or a continuous scaling factor for something like amount of reverbation. In the latter case, it is often convenient to use not the current altitude but rather the lowest altitude so far, so the violinist can dip and return to a more comfortable altitude; raising the violin above normal altitude resets the running minimum.

This mapping from position to timbre begins with a discrete pointwise map from $\mathbf{R}^2$, the space of (latitude, longitude) to $\mathbf{R}^n$, the space of filter parameters. This pointwise map is then extended to a continuous map on all of $\mathbf{R}^2$ by means of simplicial interpolation, as follows. First the set of points in $\mathbf{R}^2$ is triangulated to form a simplicial complex (triangular mesh). This induces a corresponding simplicial complex in $\mathbf{R}^n$. Finally, a simplicial mapping is defined between the two complexes, identifying corresponding simplices (triangles), and identifying points in such simplices with equal barycentric coordinates [4].

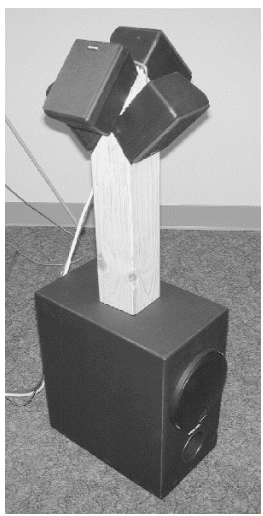

Figure 2. Quasi-spherical loudspeaker configuration for the eviolin.

We compute the approximate position of the violin's bow by subtracting bow position from violin position (vector subtraction). Bow speed is computed by subtracting this value from the value measured a moment earlier.

**LOUDSPEAKER DESIGN**

The loudspeaker arrangement is based on the Sirocco Crossfire system, intended for computer video games. It consists of a five-channel amplifier, a subwoofer, and four small "satellite" speakers (figure 2). The satellites are epoxied together to fire outwards from a point, in tetrahedral configuration; this loudspeaker-molecule sits on a 30 cm wooden column above the subwoofer. We find this to be an inexpensive yet effective version of the spherical loudspeakers used by Trueman [6].

**ACKNOWLEDGMENTS**

Michael J. Lang implemented the first version of the Max patch and the resonance models. Matt Wright graciously helped us with OSC debugging. We particularly thank Chad Peiper for rehearsing and performing compositions for the eviolin on numerous occasions.